\documentclass[%
 reprint,
 amsmath,amssymb,
 aps,
]{revtex4-2}

\usepackage{graphicx}% Include figure files
\usepackage{dcolumn}% Align table columns on decimal point
\usepackage{bm}% bold math
\begin{document}

\title{Convolutional Neural Networks for Mode On-Demand High Finesse Optical Resonator Design}

\author{Denis V. Karpov}
\email{d.karpov@soton.ac.uk}
\affiliation{Optoelectronics Research Centre, University of Southampton, Southampton SO17 1BJ, United Kingdom}

\author{Sergei Kurdiumov}
\affiliation{Optoelectronics Research Centre, University of Southampton, Southampton SO17 1BJ, United Kingdom}

\author{Peter Horak}
\affiliation{Optoelectronics Research Centre, University of Southampton, Southampton SO17 1BJ, United Kingdom}

\date{\today}

\begin{abstract}
We demonstrate the use of machine learning through convolutional neural networks to solve inverse design problems of optical resonator engineering. The neural network finds a harmonic modulation of a spherical mirror to generate a resonator mode with a given target topology ("mode on-demand"). The procedure allows us to optimize the shape of mirrors to achieve a significantly enhanced coupling strength and cooperativity between a resonator photon and a quantum emitter located at the center of the resonator. In a second example, a double-peak mode is designed which would enhance the interaction between two quantum emitters, e.g., for quantum information processing.
\end{abstract}

\maketitle

\section{Introduction}

Machine learning methods are widely used for various applications of modern optics, especially for inverse design problems \cite{nn1,nn2,nn3}. Methods of forward design, where based on a given optical device geometry and boundary conditions researchers can find the optical field and its properties, are highly developed and various numerical, semi-analytical, and analytical methods are well known. On the other hand, not often is it possible to devise geometries to generate a given target optical field using forward design methods. In this case deep machine learning methods and in particular neural networks can play the role of universal predictors and ultra-precise interpolators for various types of engineering. For example, optical power beam splitters and multiport devices for arbitrary transmission matrices have been designed with these methods \cite{bs1,bs2}.

The same approach can be applied to the design of optical resonators for cavity quantum electrodynamics and quantum technology. For example, for various applications in these areas strong coupling of a quantum emitter to photons in a resonator is required and simultaneously a long photon lifetime in this resonator is critically important to reach the so-called good cavity limit \cite{kuhn} where the coherent cavity coupling exceeds the decay rates. At the same time there often are geometrical restrictions, e.g., to allow for access from the side for loading, state initialisation, optical cooling \cite{cooling}, or trapping fields \cite{prep}. Other applications require strong photon coupling to \textit{two} quantum emitters for 2-qubit quantum gates \cite{qi1,qi2}, such as for a controlled-not (CNOT) operation between two qubits stored in trapped ions \cite{Pellizari}. In this case the performance could be significantly improved with an optimized cavity mode that exhibits two peaks at the positions of the two ions. 

There is thus a need for a universal method to design and engineer cavities used in quantum information applications \cite{comp} which should combine strong photon-particle coupling and low cavity losses by achieving a target field topology (e.g., one or several field maxima in the center and low mode divergence) while also taking into account constraints on geometry, materials, and fabrication. Traditionally, only cavities with spherical mirrors supporting Laguerre-Gaussian modes have been considered for these applications, but they provide only a limited design space that may not fulfill the requirements of specific applications. Allowing for non-spherical cavity mirrors provides a much wider range of design options and can generate cavity eigenmodes that have superior properties compared to standard Laguerre-Gaussian modes \cite{paper1,paper2}. However, the large design space then makes inverse design problems very difficult to solve, which we overcome by using a convolutional neural network (CNN) approach here.

For most optical device inverse designs, e.g., for beam splitters, the relevant quantity to be optimized is the electromagnetic field  \cite{bs1,bs2}. Optimizing an optical resonator is more complicated since a resonator supports many modes, each one of which could be a potential solution. Furthermore, it is generally not only the electric field of the mode that is of interest but also the mode decay rate must be taken into account; for example, a mode with high field intensity can have very high losses and thus be useless for real applications. In this paper we demonstrate an approach based on a coordinate-dependent cooperativity $C$, a measure that includes the ratio of the coherent coupling strength of the quantum emitter to the cavity photon (i.e., the single-photon field strength) over the incoherent cavity loss rate. Our CNN finds cavity geometries that optimize the cooperativity $C$ and that can be achieved realistically by various modern fabrication methods such as a fiber-optic microcavities \cite{Pellizzari1995,Cirac1997,Kimble2008,Monroe2013}, ion beam etched dielectric resonators \cite{fib}, and micro-assembled structures\cite{revResonators}.

This paper is organized as follows. First, in Sec.\ \ref{sec:statement}, we describe our optical resonator model and introduce the parameters for optimization. In Sec.\ \ref{sec:algo} we present a general overview of the optimization algorithm which consists of data generation, CNN training and verification. In Sec.\ \ref{sec:cnn} more details of the neural network topology and computational resources are discussed. The CNN predictions of optimized cavity mirror geometries and their verification is presented in Sec.\ \ref{sec:results}. Finally we summarize our results and conclude in Sec.\ \ref{sec:conclusion}.

%%%%%%%%%%%%%%%%%%%%%%%%%%%%%%

\section{Optical resonator coupling to a quantum emitter}\label{sec:statement}

\begin{figure}[tbp]
  \centering
  \includegraphics[height=3cm]{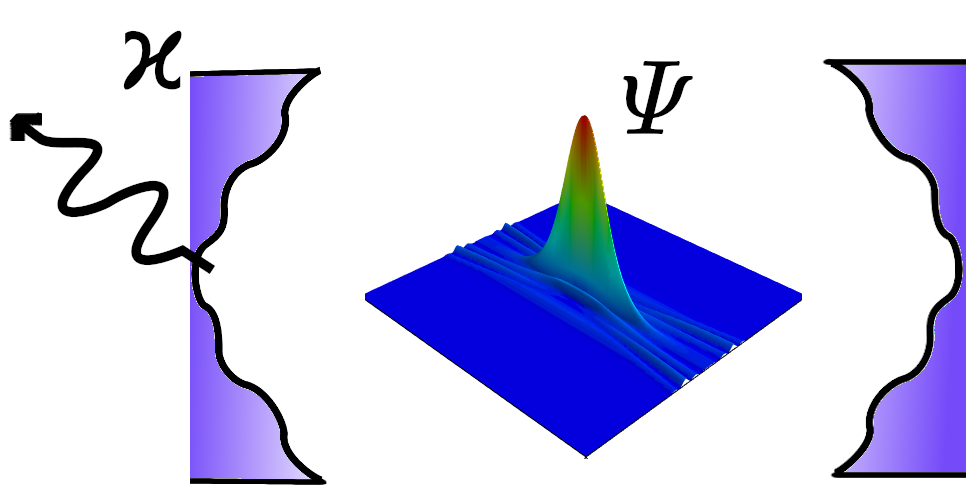}
  \caption{Schematic cavity representation. The shape of the mirrors should be optimized to generate a target cavity field mode $\Psi$ while keeping the losses $\kappa$ small.}
  \label{fig:schematic}
\end{figure}

We consider a Fabry-Perot optical cavity consisting of two mirrors whose geometry we wish to optimize for applications in quantum science and technology. A schematic of this is shown in Fig.\ \ref{fig:schematic}.

For a given cavity geometry, the cavity modes can be calculated using standard methods, such as Gaussian beam resonator theory \cite{yariv}, ray transfer matrix approaches \cite{trans1, trans2}, or finite element methods. The electric field of a chosen mode is then normalized to a single photon field $E(\textbf{r})$, i.e., the total energy contained in the electromagnetic field inside the cavity is normalized to a single photon energy $\hbar \omega$, where $\omega$ is the angular frequency. The interaction energy between a dipole emitter at position $\textbf{r}$ and a cavity photon is given by $\hbar g(\textbf{r})=\mu E(\textbf{r})$ where $g$ is called the coupling constant and $\mu$ is the electric dipole of the emitter. Exploiting the relationship between the dipole moment and the spontaneous decay rate of the emitter, 
$\Gamma= \frac{\omega^3 \mu^2}   {3\pi\varepsilon_{0}\hbar c^3}$, we can write the coupling constant as \cite{Vucovich}
\begin{equation}
g(\textbf{r})=\sqrt{\frac{3\lambda^2c\Gamma}{4\pi V_{\Psi} }}\Psi(\textbf{r}),\
\Psi(\textbf{r}) = \frac{E(\textbf{r})}{|E(\textbf{r}_m)|}
\label{eq:coupl1}
\end{equation}
where $\lambda$ is the wavelength corresponding to angular frequency $\omega$, $c$ is the vacuum speed of light, $\Psi$ is the dimensionless cavity field normalized to its maximum at position $\textbf{r}_m$ and the mode volume is defined by
\begin{equation}
V_{\Psi} = \int_{V_{cavity}} |\Psi(\textbf{r})|^2 dV
\label{eq:volume}
\end{equation}
where the integral is over the geometric volume of the cavity.

The other important quantity of optical cavities for quantum applications is the cavity decay rate $\kappa$, i.e., the rate at which a photon is lost from the cavity mode. We distinguish two contributions to the cavity loss rate: the loss per photon round trip by transmission through the mirrors or by absorption within the mirror $D_{mir}$ (these are typically of the order of $10^{-5}$ - $10^{-3}$), and the loss per round trip through so-called clipping losses $D_{clip}$ where the mode on the mirror is larger than the mirror diameter, i.e., a fraction of the photon field misses the mirror completely on reflection. The cavity decay rate is then
\begin{equation}
    \kappa = (D_{clip}+D_{mir})\frac{c}{2L}
\label{eq:kappa}
\end{equation}
where $L$ is the length of the cavity. An alternative way to express the loss rate is via the cavity finesse $F$ defined as
\begin{equation}
F = \frac{2\pi}{D_{clip} + D_{mir}} = \frac{c\pi}{L} \frac{1}{\kappa}
\label{eq:finesse}
\end{equation}

To achieve strong particle-cavity coupling, the coherent coupling rate $g(\textbf{r})$ between the particle at position $\textbf{r}$ and the cavity must be larger than the strengths of any incoherent processes, i.e., energy exchange between the particle and the cavity must occur on time scales before the photon leaks out of the cavity or is incoherently scattered by the atom. Therefore, the cooperativity parameter defined as
\begin{equation}
C(\textbf{r}) = \frac{g(\textbf{r})^2}{\kappa \Gamma}  = \frac{3\lambda^2c}{4\pi \kappa V_{\psi}} |\Psi(\textbf{r})|^2
\label{eq:coop}
\end{equation}
must be larger than one. The cooperativity therefore not only depends on the single-photon electric field strength at the position of the atom, but it also includes the round trip losses $D_{clip}$ and $D_{mir}$. 

Our approach is to modify the shape of the cavity mirrors based on a spherical ``reference'' shape. It is therefore useful to compare the cooperitivity achieved with our designs to the maximum cooperativity obtained at the center of this reference cavity. For two identical spherical mirrors with radius of curvature $R$ \cite{Hunger2010} the fundamental cavity mode is given by a Gaussian beam and the maximum cooperativity becomes
\begin{equation}
C_0 = \frac{6\lambda F}{\sqrt{2RL-L^2}} 
= \frac{6\lambda}{\sqrt{2RL-L^2}} \frac{2\pi}{D_{mir}} 
\label{eq:coop2}
\end{equation}
where we assumed no clipping losses for this reference cavity mode.

For quantum optics and quantum technology applications, the cooperativity $C$ is the quantity that we wish to optimize by a convolutional neural network approach in the following sections. Our CNN will thus be trained on a set of functions $C(\textbf{r})$ or equivalently, once a reference spherical mirror shape is chosen, on the relative enhancement $C(\textbf{r})/C_0$, which constitutes a fundamental difference to previous CNN-based optimization of optical devices which only take into account the electric field itself, e.g., in the case of integrated optical beamsplitters \cite{bs1,bs2}.

%%%%%%%%%%%%%%%%%%%%%%%%%%%%%%%%%%%%%%%%%%%%%%%

\section{Algorithm}\label{sec:algo}

\begin{figure*}[tbp]
  \centering
  \includegraphics[width=0.99\textwidth]{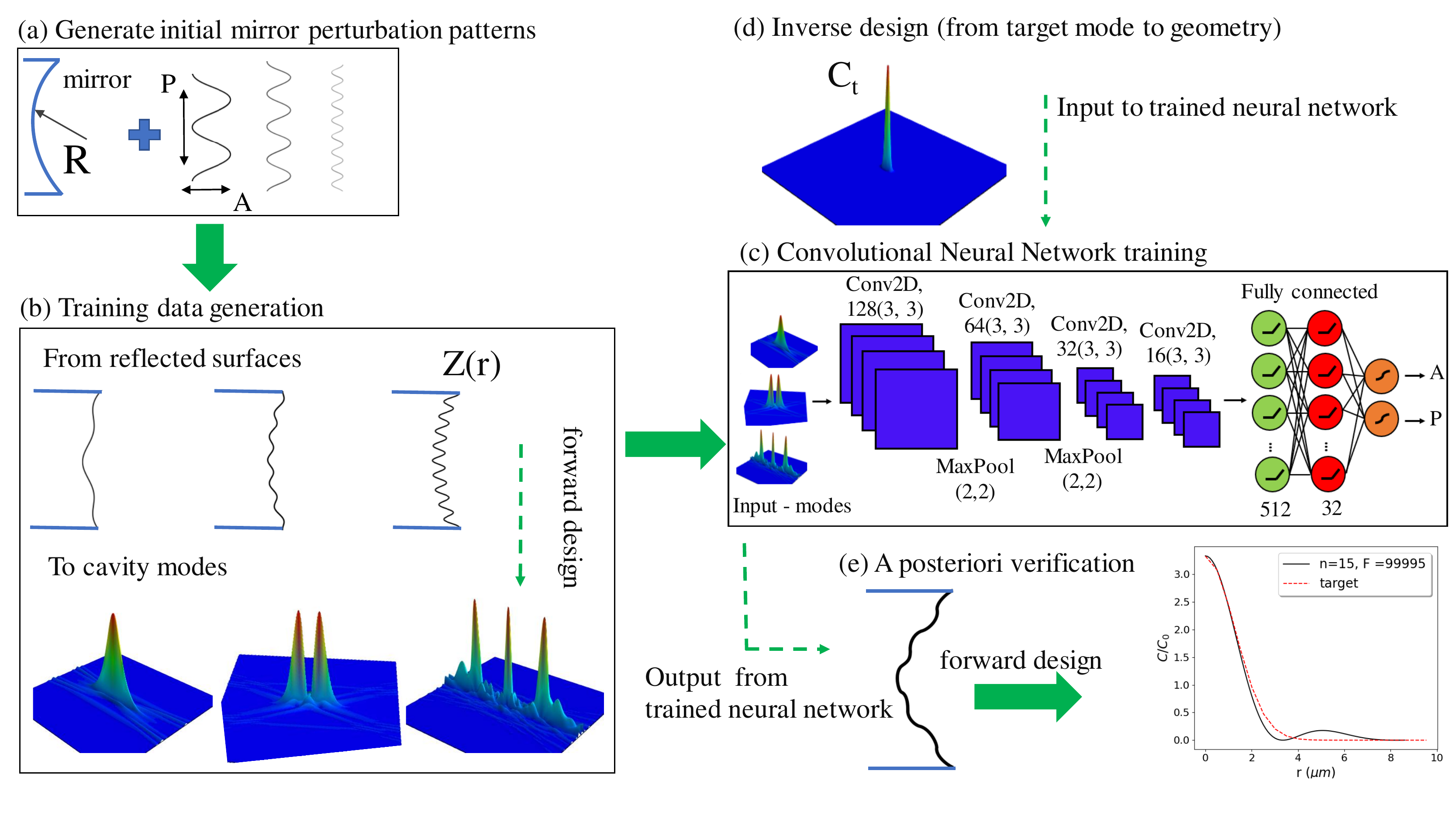}
  \caption{Overview of the optimization algorithm including data generation (a-b), machine learning (c), inverse design prediction (d) and verification (e). (b) shows modes with various topologies which exist within the design space. (d) represents a sample target cooperativity distribution (i.e., a peak of high intensity and low loss) as input argument for the trained CNN. (e) compares the target mode from (d) with the mode calculated explicitly from the CNN-generated mirror geometry.}
  \label{fig:algo}
\end{figure*}

We consider cavities with cylindrical rotational symmetry consisting of two identical mirrors with a spherical profile perturbed by a harmonic modulation given by
\begin{equation}
Z(r)=\frac{r^2}{2R} + A\cos(r/P).
\label{eq:deviations}
\end{equation}
By limiting the ranges of the period $P$ and the amplitude of the perturbation $A$, we can choose parameters which are achievable by modern fabrication tools, see Sec.\ \ref{fab}. However, we note that our approach would work for any parameterization of the mirror surface. Eq.\ (\ref{eq:deviations}) is valid within the paraxial approximation, i.e., we approximated the spherical profile by a parabolic one.

To generate our data training set we need a method to find the cavity eigenmodes for a given mirror profile $Z(r)$. This \textit{mode solver} will be treated as a black box function below. Any numerical or semi-analytical method can be used as mode solver; for the numerical calculations in this paper we employ a mode mixing method (MMM) \cite{Kleckner2010,Nina}. The advantages of the MMM, compared to, e.g., finite elements or finite differences methods, are high computation speed, low RAM and CPU consumption and the absence of sharp corner singularities \cite{sing1,sing2}. Depending on the complexity of the geometry one simulation on a single CPU core takes no more than 20 seconds, which allowed us to generate our CNN training data of 85000 samples on 240 CPU cores on an HPC cluster (2.0 GHz Intel Skylake processors) in about 2 hours of computation time. Choosing a computationally effective method for the mode solver is critical for our approach.

Once the mode solver has calculated the modes $\Psi_i$ and their finesse $F_i$ for a given $Z(r)$ we proceed by selecting one mode based on the target topology, e.g., a mode with a single peak, two peaks, etc. The spatially dependent cooperativity $C(\textbf{r})$, Eq.\ (\ref{eq:coop}), of this selected mode is then added to the training database. We emphasize that this selection process is very important to generate a well balanced training set; different training sets are required for different target mode topologies.

An overview of our algorithm is presented in Fig.\ \ref{fig:algo}. The first step is the selection of a perturbation pattern imposed on the spherical mirror within some range of parameters, Fig.\ \ref{fig:algo}(a). In our case we choose harmonic perturbations with a flat distribution of periods $P$ in the range 5 $\mu m$ to 30 $\mu m$ and amplitudes $A$ in the range 0.1 to 0.6 $\mu m$. The second step is to use these geometries as arguments for the mode solver, calculate the modes, and select the most appropriate mode for the training data set as outlined above, Fig.\ \ref{fig:algo}(b). The third step, Fig.\ \ref{fig:algo}(c), is to train the CNN on the generated data set. This is described in more detail below in Sec.\ \ref{sec:cnn}.
%Here we tune the internal network structure, conduct test optimisation in according of computation resources usage, decide which cooperativity resolution would be satisfied and choose crop frame, more details about CNN topology in the next Sec.\ \ref{sec:cnn}. 
The fourth step is testing the trained CNN on data which were not used in the training (more details below) and usage of the trained CNN as a predictor for inverse mirror design: a target cooperativity distribution is given to the CNN and it predicts the required mirror modulation parameters $P$ and $A$, Fig.\ \ref{fig:algo}(d). At the final, fifth step we put this predicted mirror geometry into the mode solver and verify the result of the mode field, Fig.\ \ref{fig:algo}(e), as discussed in more detail in Sec.\ \ref{sec:results}.

%%%%%%%%%%%%%%%%%%%%%%%%%%%%%%%%%%%%%%%%%%%%%

\section{Convolutional Neural Network}\label{sec:cnn}

\subsection{Introduction to neural network approach}\label{sec:cnn.intro}

Deep learning methods have recently been demonstrated as a powerful tool for solving physical and in particular optical problems, for example, for deeply subwavelength optical imaging \cite{im1,im2,im3}, analysis of scatterometry data \cite{scat1,scat2}, enhanced resolution of SEM images \cite{sem}, MRI image analysis \cite{mri1,mri2}, inverse design of optical components \cite{bs1, bs2} and many other image processing applications, revolutionising their future development.  Deep learning methods are machine learning methods employing multilayer (3 and more layers) neural networks where subsequent layers extract finer level features from the raw input data. Deep neural networks automatically find the correct mathematical transformations to convert the input data to outputs results without requiring prior knowledge of the linear or nonlinear correlations. Convolutional neural networks (CNN) \cite{cnn}, long short-term memory (LSTM) \cite{lstm}, variational autoencoders (VAE) \cite{vae}, and generative adversarial networks (GAN) are examples of deep learning algorithms used for physical tasks \cite{gan}.

\subsection{Specific network topology}\label{sec:cnn.specific}

In our paper we present the application of a deep convolutional neural network for predicting the required parameters of the cavity (period and amplitude of the sinusoidal mirror profile modulation) from a target cooperativity profile $C(\textbf{r})$. The forward design mode solver outputs a set of modes for one geometry. Depending on the application we then select the mode of highest cooperativity, highest finesse, two-peaked cooperativity etc. from the full set of modes corresponding to each geometry. Each mode is represented as a field map in a longitudinal and radial cross-section of the cavity of 500 $\mu$m length in z-direction and 400 $\mu$m in diameter. Each field map consists of $200\times 200$ points. Exploiting the fact that all modes are radially and center-symmetric, we feed the neural network with only a quarter of the field maps ($100\times 100$) in order to save computer memory.

The neural network we use in our studies consists of four convolutional layers with 3x3 kernels with max-pool layers inserted between the first three convolutional layers \cite{NN_book1}. The topology of the neural network is shown in Fig.\ \ref{fig:algo}(c). All convolutional layers are activated by rectified linear unit (ReLU) functions. The final convolutional layer is connected to three fully connected layers with 512, 32 and two neurons, respectively. The hidden fully connected layers have ReLU activation functions, and the output layer is activated by sigmoid functions \cite{activation}. These fully-connected layers are needed for linking the feature representations from the convolutional layers with the output labels. This is a typical structure of a convolutional neural network which allows for automatic feature extraction for solving the inverse problem of finding the design of a device from the generated field map. The network was trained with the Adam stochastic optimization method with a learning rate of 0.001. The mean square error (MSE) was used as the loss function.

The neural network takes the field map as an input, outputting the period and amplitude both normalized to their maximum values in the whole dataset (i.e., between 0 and 1). 80\% of the whole dataset was used for training, 10\% for validation and the last 10\% for testing.

\subsection{Technical details}\label{sec:cnn.details}

We use a self-made mode solver based on the MMM \cite{Kleckner2010, Nina} to generate the training data. The data contains 85000 numerical simulations, where each sample is a full set of modes with spatial field distribution and finesse. We use the python programming language with libraries numpy for computation and mpi4py for distribution of simulations over an HPC cluster. 

The CNN training software was also written in the python language. The open source machine learning framework Tensorflow was used for the deep neural networks and training took place on an Nvidia GPU supplied with CUDA.

%%%%%%%%%%%%%%%%%%%%%%%%%%%%%%%%

\section{Results and discussion}\label{sec:results}

\subsection{Mode prediction and verification}

\begin{figure}[!tbp]
{(a) \hspace{5cm}}\\
\includegraphics[width=0.5\textwidth]{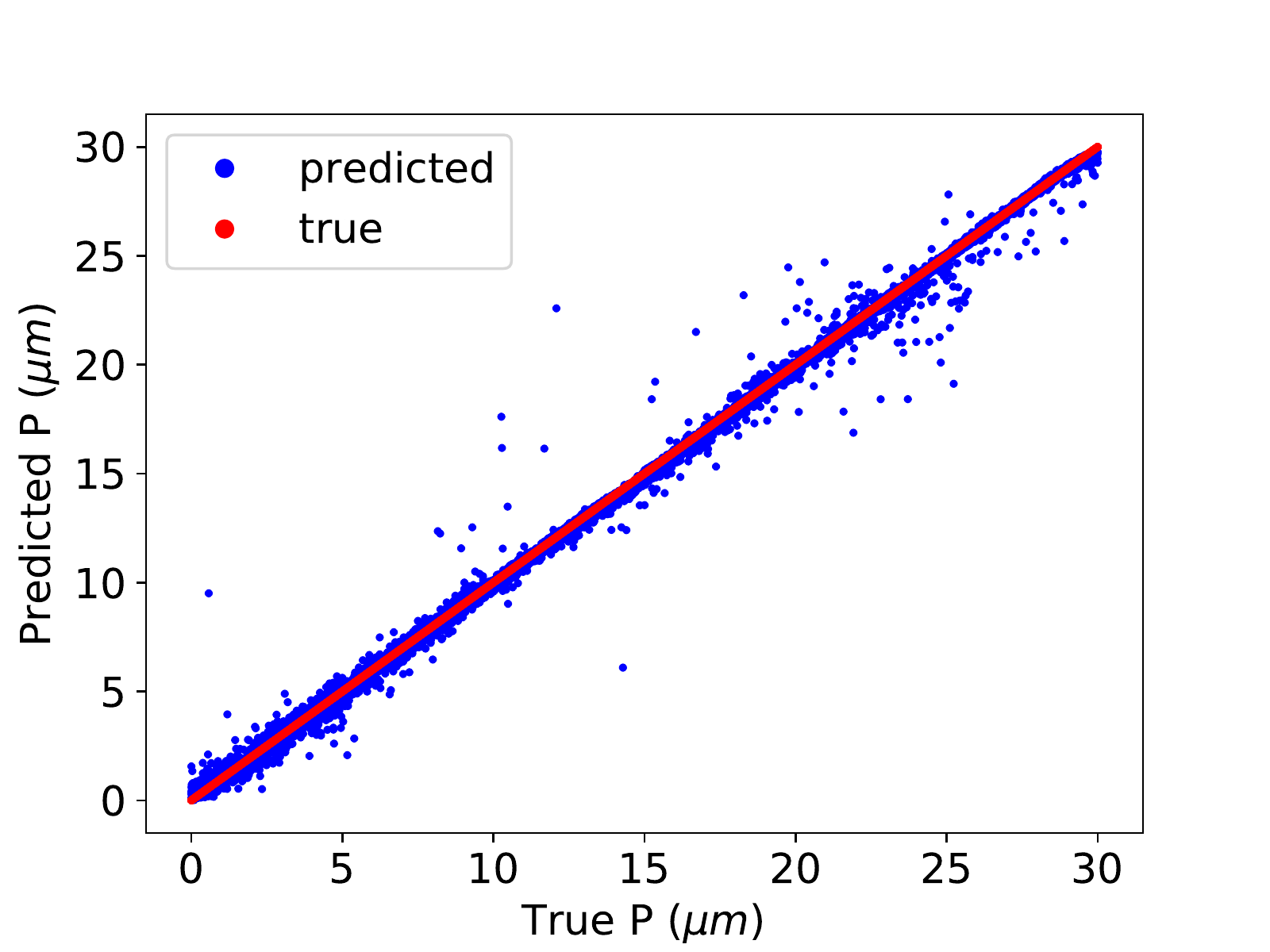}
\vspace{0.5cm}
{(b) \hspace{5cm}}\\
\includegraphics[width=0.5\textwidth]{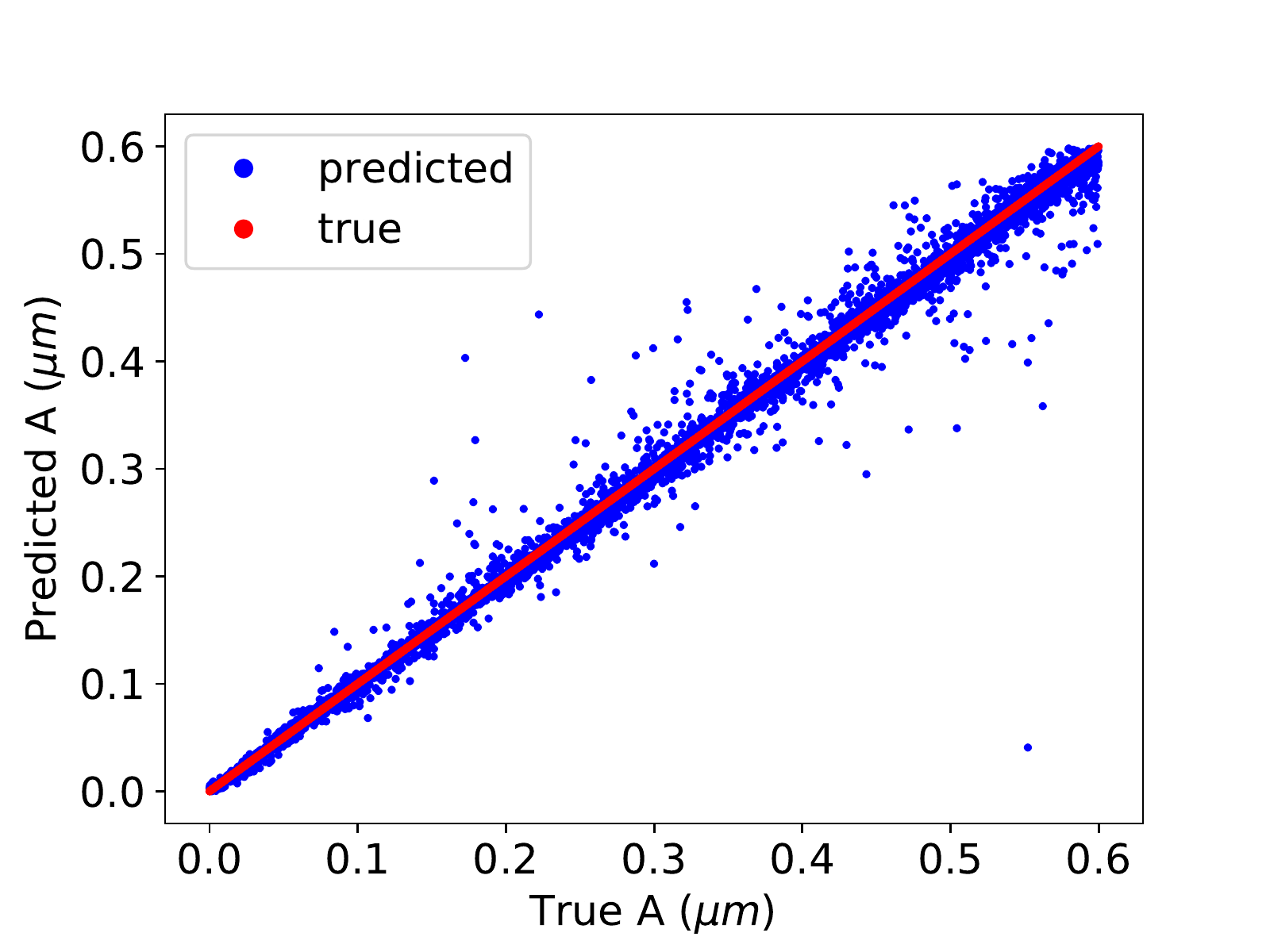}
  \caption{Correlation coefficient between the true parameters of the harmonic mirror modulations (red dots) and the values predicted by the trained network (blue dots) for (a) period $P$ and (b) amplitude $A$. The correlation coefficients are 0.998826 and 0.996890 for period and amplitude, respectively, after training with 90\% (10\% of that was used for validation) and for a test set of 10\% out of $N = 85000$ data sets.}
\label{fig:learning}
\end{figure}

For the examples discussed here we choose a reference spherical cavity of length $L=500~\mu$m, mirror radius of curvature $R=400~\mu$m, and mirror diameter $200~\mu$m. Harmonic modulations of the mirror profiles are chosen with random period $P$ and amplitude $A$ as discussed in Sec.\ \ref{sec:algo}. 

The results of the neural network training are shown in Fig.\ \ref{fig:learning}. This plot compares true parameters (red line) with CNN predictions (blue dots). This curve is based on a 10\% subset of the forward design data set of 85000 samples which were separated from the training set and have not been used in training or validation (made after every epoch by CNN). From the figure we see that the correlation coefficient between the true and retrieved values of both amplitude and period are over 99\%, which demonstrates a high quality of the retrieval process. 

\begin{figure}[tbp]
  \centering
  \includegraphics[width=0.5\textwidth]{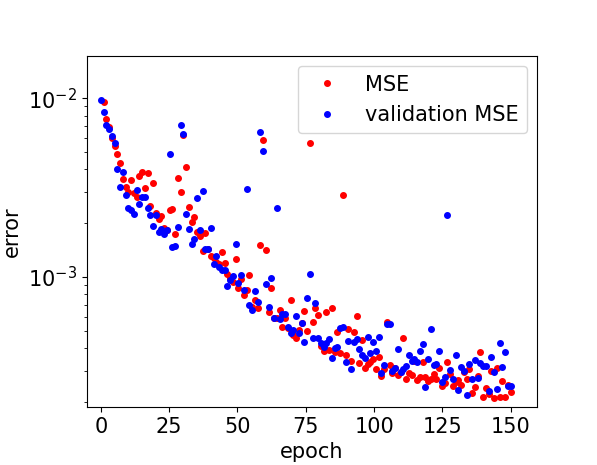}
  \caption{Learning curves for CNN training: mean square error of the training set (red dots) and of the validation set (blue dots) during the learning procedure within 150 epochs.}
  \label{fig:epoch}
\end{figure}

Figure \ref{fig:epoch} presents the training convergence curves (learning curves) where training MSE and validation MSE versus training epoch show a common, smooth convergence. This confirms that the neural network does not overfit in the process of training.

\begin{figure}[!tbp]
{(a) \hspace{5cm}}\\
\includegraphics[width=0.4\textwidth]{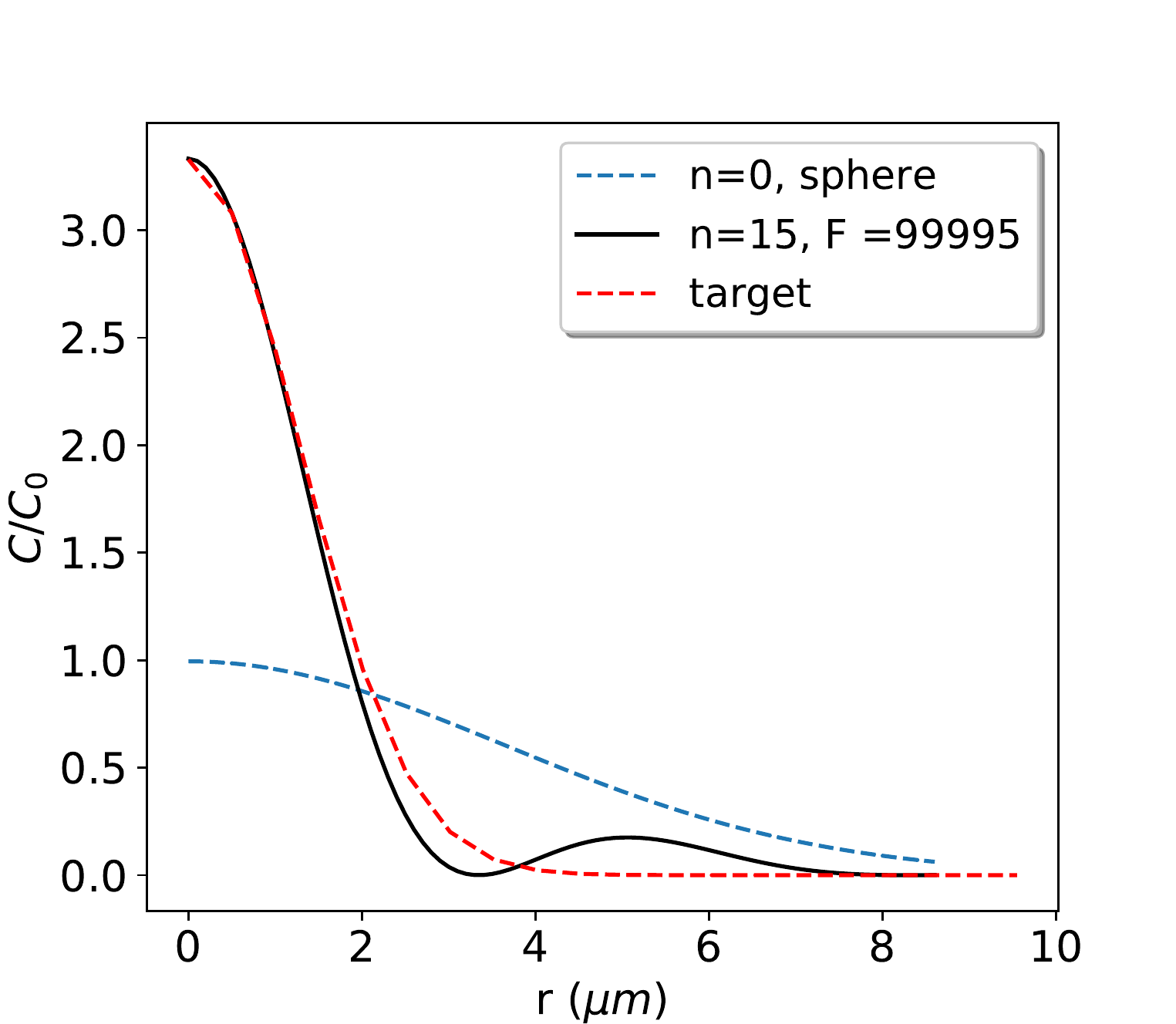}\\
\vspace{0.5cm}
{(b) \hspace{5cm}}\\
\includegraphics[width=0.4\textwidth]{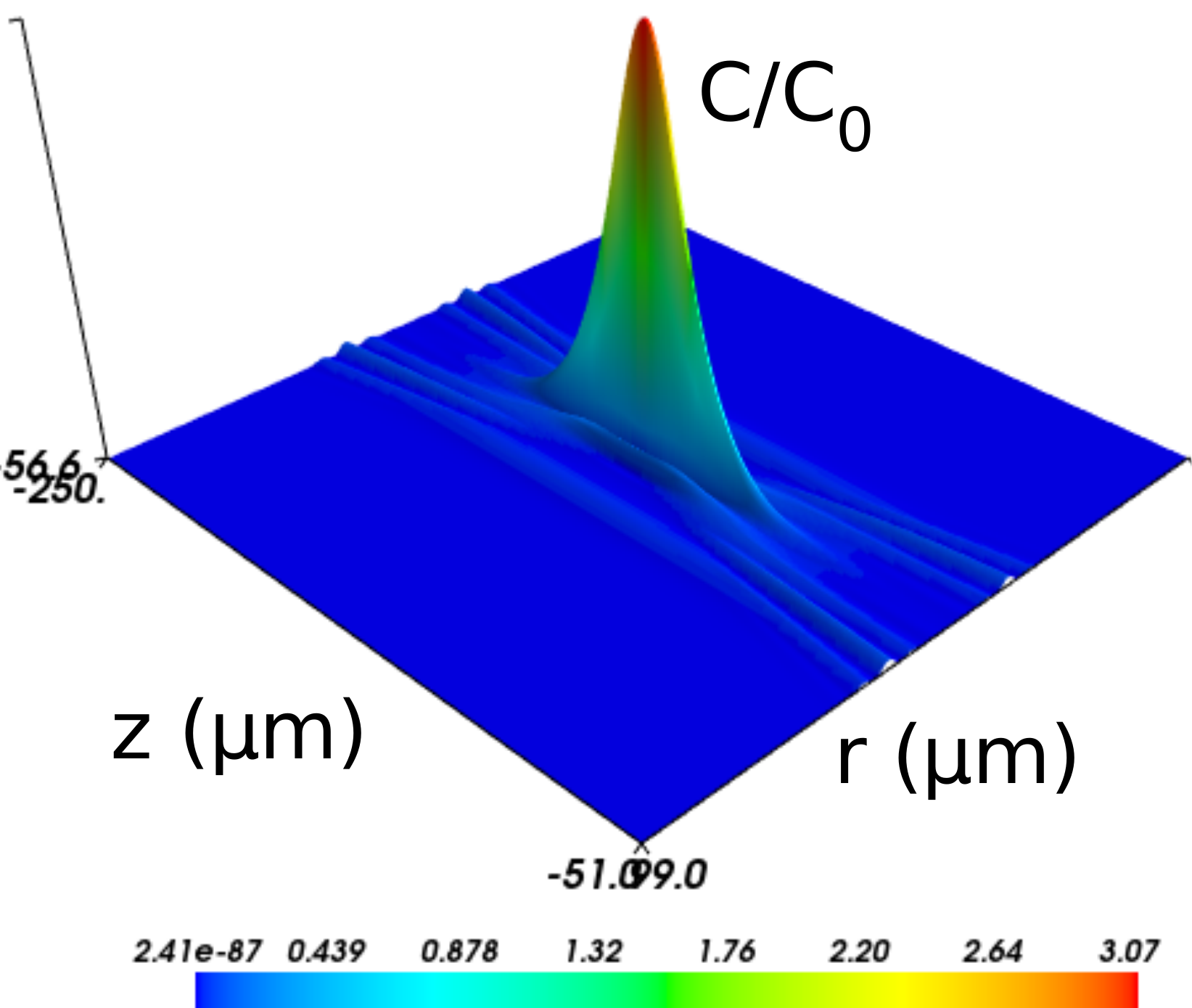}
  \caption{Demonstration of CNN inverse design. (a) shows the radial cross section of the target mode (red dashed curve), the numerically verified predicted mode (black solid), and the fundamental Gaussian mode of the spherical cavity of the same size (blue curve). $n$ is the mode number. (b) shows the predicted mode in cylindrical coordinates.}
\label{fig:predict}
\end{figure}

Next we use the trained CNN to predict the mirror modulation parameters for a target mode profile. This target mode in Fig.\ \ref{fig:predict} was chosen to provide a desired cooperativity enhancement at the cavity center and a specific mode shape. Then we use the predicted mirror profile in our mode solver to confirm the generated mode field. Figure \ref{fig:predict}(a) compares the target mode (red dashed curve) and the predicted mode (black solid) in a radial cross section. The blue curve corresponds to the fundamental Gaussian mode of the spherical cavity of the same size, confirming that we achieved an enhancement of cooperativity by a factor of three. Figure \ref{fig:predict}(b) shows the predicted mode profile in cylindrical coordinates.

\begin{figure}[!tbp]
{(a) \hspace{5cm}}\\
\includegraphics[width=0.4\textwidth]{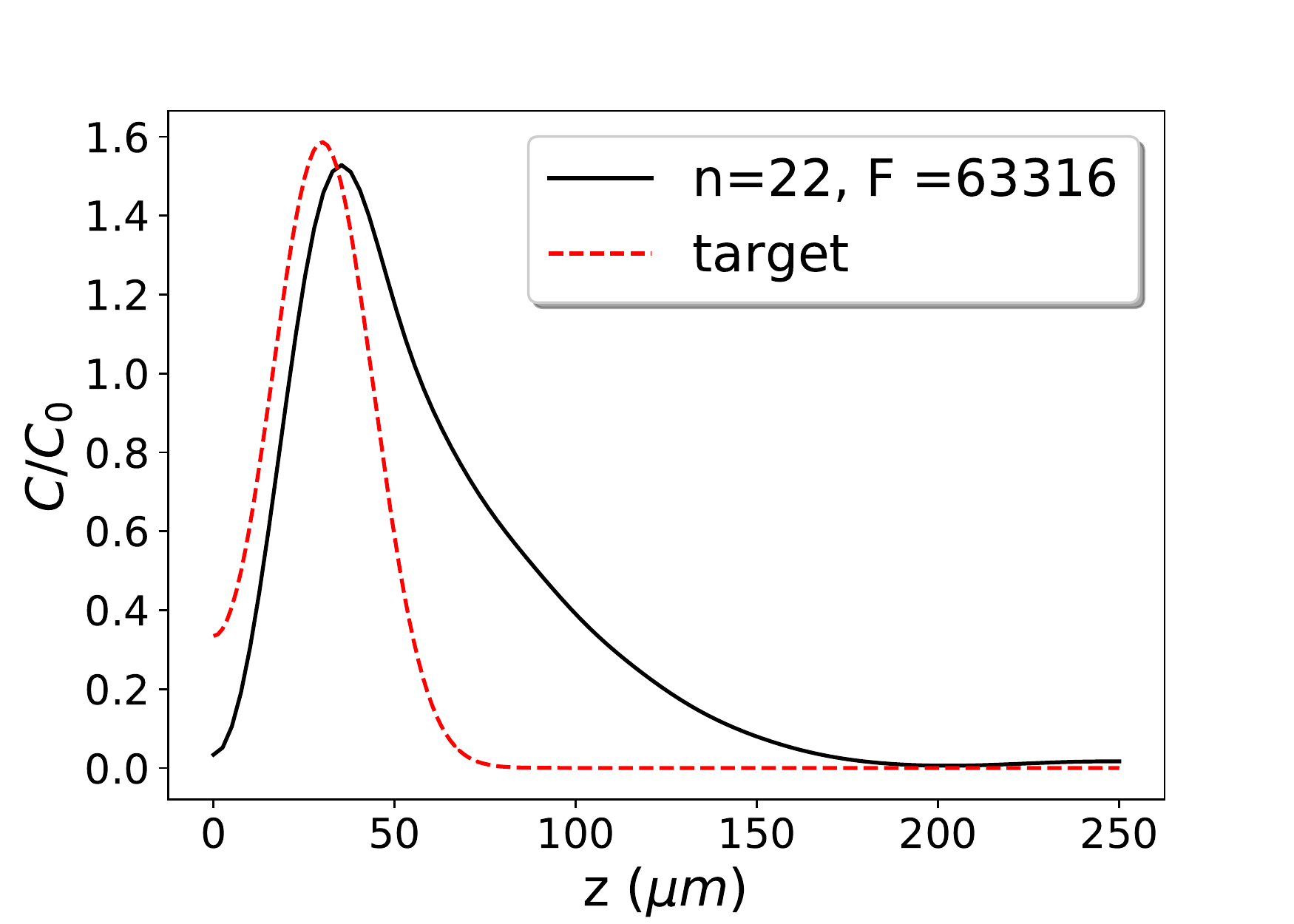}\\
\vspace{0.5cm}
{(b) \hspace{5cm}}\\
\includegraphics[width=0.4\textwidth]{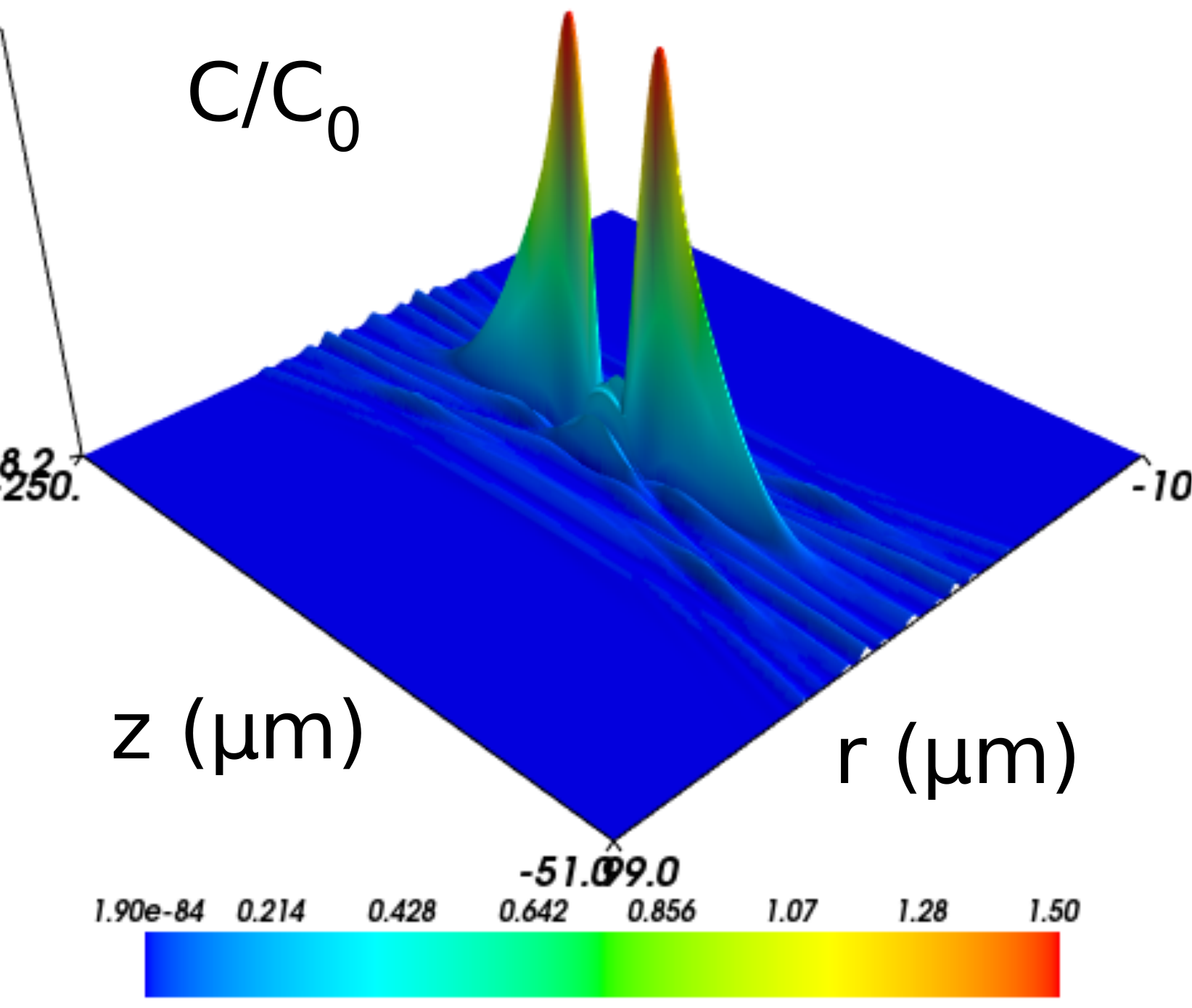}
  \caption{Demonstration of CNN inverse design. (a) shows the axial cross section of the target mode (red dashed curve) and the numerically verified predicted mode (black solid). (b) shows the predicted mode in cylindrical coordinates confirming the two-peak structure.}
\label{fig:double}
\end{figure}

In a second example, Fig.\ \ref{fig:double}, we target a mode comprising two maxima. We define the target as the sum of two Gaussian profiles shifted apart along the z-axis. The cross section of the target and the predicted modes along the z-axis are shown in Fig.\ \ref{fig:double}(a). The shape of the  predicted mode is slightly different from the target mode, which respects the fact that not every mode topology can be supported by our family of mirror shapes, i.e., a double-peak Gaussian mode is not necessarily an eigenmode of a spherical mirror with any single harmonic modulation. However, in this case the CNN finds the closest mode to the target mode and Fig.\ \ref{fig:double}(b) shows that the main feature of the target mode of two peaks on the z-axis is well reproduced.

The capability of the CNN to predict modes with a desired topology depends on the training set composition, especially on the balance between desired modes and other modes contained in the set. We should emphasize that the training sets for the one-peak mode (Fig.\ \ref{fig:predict}) and the two-peak mode (Fig.\ \ref{fig:double}) are very different. For every mirror geometry our mode solver generates a set of modes; the exact number depends on the desired precision and in our numerical simulations was set to 30 modes. However, only one mode of this set is included in the training data and all other modes are ignored. An essential part of our method is therefore to define and computationally implement this selection rule. Depending on the selection rule we create different training sets for different target mode topologies.

\subsection{Errors and limitations of the design method}

\begin{figure}[!tbp]
{(a) \hspace{5cm}}\\
\includegraphics[width=0.4\textwidth]{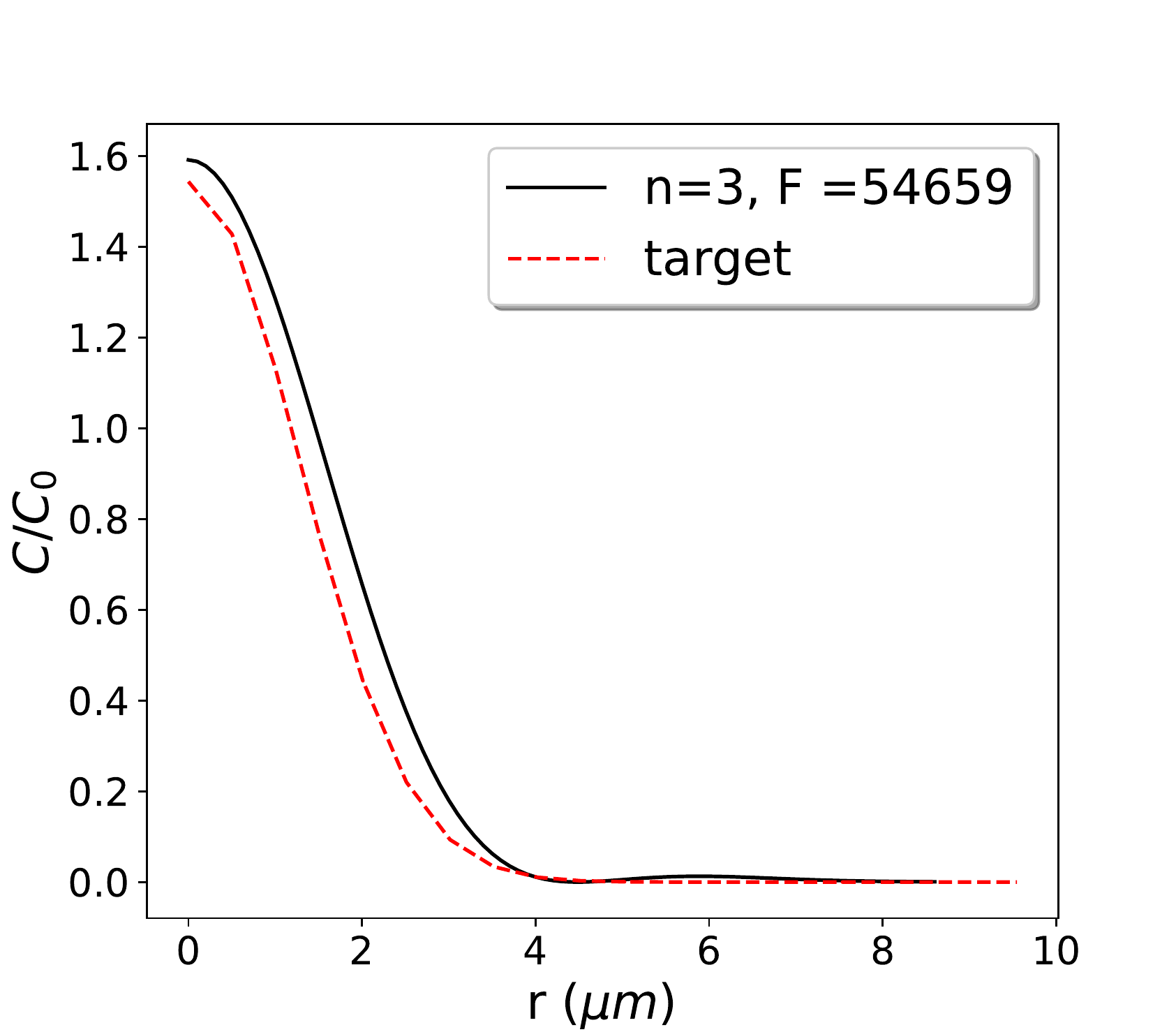}\\
\vspace{0.5cm}
{(b) \hspace{5cm}}\\
\includegraphics[width=0.4\textwidth]{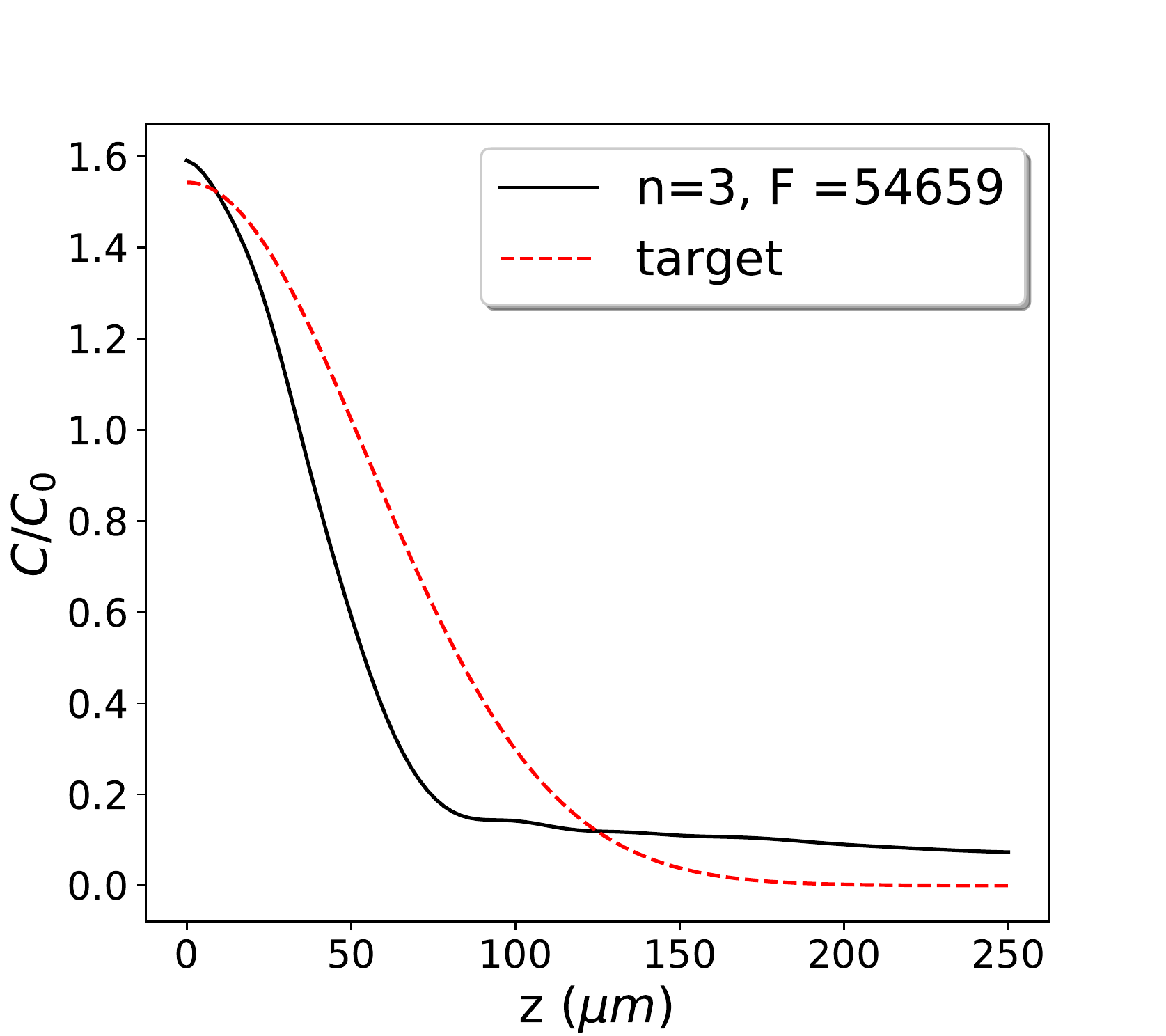}
  \caption{Attempted inverse design of a target mode (red dashed curve) that is not within the topologies of the training set. The numerically verified mode of the CNN prediction (black solid curve) matches in the radial direction (a) but deviates in the longitudinal direction (b).}
\label{fig:verif}
\end{figure}

The essential limitation of our method is given by the nature of the neural network learning process. The neural network cannot predict modes which are very different from the modes in the training set because the CNN is effectively an interpolator. If we set a target mode which cannot exist within the chosen parameter range of the training set, for example a target mode which can only exist for a different type of mirror geometry, then the CNN predicts the closest mode to the target one. In  Fig.\ \ref{fig:verif} we present such a case: here the target mode is a Gaussian shape that decays to zero both in the radial and the longitudinal direction. Such a mode is clearly unphysical as the light must always be bouncing between the cavity mirrors and thus cannot decay to zero in the longitudinal direction. Thus, the CNN finds a mode that matches the target mode accurately in the radial direction, but the mode deviates from the target in the axial direction. This is typical for this kind of methods when the target field topology is not within the topologies in the training set.

\subsection{Fabrication considerations}\label{fab}

The mirror geometries investigated here require the fabrication of a harmonic modulation with scales of tens of micrometers in the radial direction and hundreds of nanometers in the axial direction on a spherical mirror profile. These scales are defined by the ranges of parameters which were used for the generation of the training data. Since a CNN is an interpolator but not an extrapolator, any predictions of mirror geometries beyond the training range of parameters will be inaccurate. Mirror profiles with our specifications can be created by a number of fabrication techniques, such as mirror shaping by laser machining in silica glass \cite{laserFab1, laserFab2, laserFab3} or by focused ion beam milling \cite{fib}. Pulsed CO$_2$ lasers can be used for thermal evaporation of surface material \cite{Hunger2010} generating surfaces with extremely low roughness. Also modern micro-machining tools can provide sufficient precision \cite{revResonators, mech1, mech2, mech3}.

%%%%%%%%%%%%%%%%%%%%%%%%%%%%%%%%%%%%%%%%

\section{Conclusions}\label{sec:conclusion}

We have demonstrated an approach based on convolutional neural networks for the design of high-finesse optical resonators which support a desired electrical mode field and, crucially, simultaneously limit the mode decay rate. This allows us to design target fields with, for example, one or two field maxima on the cavity longitudinal axis where significant enhancement of the cooperativity between quantum emitters coupled to cavity photons is achieved. Such cavity designs are  crucial for various quantum optics and quantum technology applications. The proposed ``mode on-demand technique'' can be extended to many types of mode topologies by implementing appropriate  mode selection rules for the training set generation.

Numerically the most demanding part of our approach is the generation of the training data set. We used an efficient mode mixing method that works in the paraxial limit of cavity optics. However, other methods can be chosen for improved accuracy or implementation of the algorithm for other inverse optical design problems. This flexibility of our CNN formalism will make it useful for many practical applications in the future. 

The data supporting this study are openly available from the University of Southampton repository \cite{dataset}.

%%%%%%%%%%%%%%%%%%%%%%%%%%%%%%%%%%%%%%%%

\section{Acknowledgments}

We acknowledge financial support by the UK Quantum Technology Program under the EPSRC Hub in Quantum Computing and Simulation (EP/T001062/1). The University of Southampton supercomputer Iridis 5 was used for numerical simulations.

%%%%%%%%%%%%%%%%%%%%%%%%%%%%%%%%%%%%%%%%%%

\end{document}